\documentstyle[12pt]{article}
\begin{document}
\thispagestyle{empty}
\begin{flushright}
FT-412-1996
\end{flushright}
\vspace*{2.5cm}
\begin{center}
{\bf UNITARITY CONSTRAINTS ON THE B AND B$^*$ FORM FACTORS}
 \end{center}
\begin{center}
{\bf FROM QCD ANALYTICITY }
\end{center}
\begin{center}{\bf AND HEAVY QUARK SPIN SYMMETRY }
\end{center}
\vskip 1.0cm
\centerline{\bf I.Caprini and C. Macesanu}
\vskip 0.5cm
\centerline {Institute for Atomic Physics, Bucharest}
\centerline {POB MG 6, R-76900 Romania}
\vskip 2.5cm
\nopagebreak\begin{abstract}
\noindent
A  method of deriving bounds on the weak meson form factors, based on
perturbative QCD, analyticity and unitarity, is generalized in 
 order to fully exploit
 heavy quark spin symmetry in the ground state $(L=0)$ doublet
of pseudoscalar $(B)$ and vector $(B^*)$ mesons. 
All the relevant form factors of these mesons are
taken into account in the unitarity sum. They are treated as
 independent functions along the timelike axis,
being related by spin symmetry only near the zero recoil point.
Heavy quark vacuum polarisation
 up to three loops in perturbative QCD and the experimental cross 
sections $\sigma(e^+e^- \rightarrow \Upsilon)$ are used as input.
We obtain  bounds on the charge radius of the elastic form factor
of the $B$ meson,
which considerably improve previous results derived in
the  same framework. 

\end{abstract}
\newpage\setcounter{page}1
\section {Introduction}
Bounds on the charge radius of the elastic
 form factor of the $B$-meson were recently derived in a number of 
papers \cite{raf}-\cite{boyd}.
 The interest in this form factor comes from the fact that it
 coincides, in the large quark mass limit, with
the renormalized Isgur-Wise function of the heavy quark effective 
theory \cite{isgur},\cite{review}. The short distance
 and finite mass corrections are in this case much smaller than
 for the flavor changing currents involved
 in the semileptonic decays of the $B$ meson into $D$ or $D^*$. 
Therefore, 
rigorous bounds on this form factor are of interest for testing various
 nonperturbative techniques applied for the calculation of the Isgur-Wise
function.

The method applied in Refs. \cite{raf}-\cite{boyd}, based on  previous
 works \cite{meiman}
starts by exploiting the same input as  the standard QCD 
sum rules, i.e. the QCD euclidian expansion of a polarization function,
 related by 
analyticity and unitarity to the physical states of interest. However,
 while in the usual formulation of the QCD
sum rules one tries, by suitable methods, to enhance the contribution of the 
low energies in the dispersion integral and saturates the unitarity sum by
the lowest lying resonances, in the approach proposed in \cite{meiman} the 
dispersion relation is written as a rigorous integral inequality for
the modulus squared of the form factors
of the physical states along the time like region. By using in addition 
the analyticity
properties of the form factors, this inequality is  shown
to  constrain  the behaviour of these functions or their derivatives near
 the zero recoil or other points of physical interest.

In refs. \cite{raf}-\cite{boyd} the method was applied to the elastic
 form factor of the
pseudoscalar $B$ meson. An attempt to exploit  heavy quark symmetry 
in the ground $B$ meson state doublet was made in 
\cite{raf}, where the $B\bar B$ and $B\bar B^*+B^*\bar B$ intermediate 
states were included in the unitarity sum, with the 
additional assumption that the
relevant form factors of the $B$ and $B^*$ mesons are identical
 along the whole unitarity cut.
 However, this is  an unjustified extension of the
 heavy quark spin symmetry, which is valid
 only near the zero recoil point.
As illustrated
in \cite{ball} by specific models, the $B$ and $B^*$
 form factors can be indeed quite
different along the time like axis, especially  near thresholds.
The problem was correctly solved in \cite{capspin}, where by means of special
 techniques
allowing the simultaneous treatment of several analytic functions 
\cite{duren},\cite{capcomp},
 the inclusion the  form factor of the $B\rightarrow B^*$ 
transition was possible within the strict heavy
quark spin symmetry hypotheses. More precisely, the elastic form factor
and the $BB^*$ form factor were treated as distinct
functions along the unitarity cut, being assumed to coincide only near the
zero recoil point. This led to a considerable improvement of
the bounds on the charge radius of the $B$ elastic form factor: 
 the range $-4.5 \leq \rho^2 \leq 6.1$, obtained in \cite{raf1}
 without
imposing spin symmetry, was narrowed  in 
\cite{capspin} to $-0.90\leq \rho^2\leq 2.60$.

However, in \cite{capspin} spin symmetry 
  was not fully
exploited, as the contribution of the $ B^*\bar B^*$ intermediate states 
in the unitarity sum was not included.
This problem is addressed in the present paper, where we treat
simultaneously all the weak form factors of the $B$ and $B^*$ mesons. 
 The quadratic expression yielded
by unitarity is written in a suitable "diagonal" form, which allows us
to apply the optimization theory  for vector-valued analytic functions
\cite{duren} \cite{capcomp}.
The different thresholds in the unitarity sum and the
subthresholds singularities of the various form factors are 
taken into account correctly.
In this way  the consequences of spin symmetry
in the ground state doublet of the $B$ mesons are exploited in an 
optimal way. The present paper
contains in addition two important improvements of the work done 
before: we use as input
the heavy quark vacuum polarization function computed in 
perturbative QCD up to three loops \cite{svz}-\cite{kuhn} and
we include in the dispersion relation 
for the polarization function  the three $\Upsilon$
 resonances with masses below 
the threshold for $B\bar B$ production (these terms were neglected in previous
works \cite{raf}-\cite{boyd}).

In the next section we present the derivation of the bounds. Section 3
contains the numerical results and our conclusions.
\section{The derivation of the bounds}
We use the standard definitions of
 the   form factors of the $B$ and $B^*$
mesons \cite{raf},\cite{ball}:
\begin{equation}\label{deff1}
< B(p')\vert V^\mu \vert B(p)>= (p+p')^\mu F(q^2)
\end{equation}
\begin{equation}\label{deff2}
< B^*(p',\epsilon )\vert V^\mu \vert B(p)>={2i\epsilon^{\mu\nu\alpha\beta}
\over m_B+m_{B^*}}\epsilon_\nu p'_\alpha p_\beta V(q^2)
\end{equation}
\begin{eqnarray}\label{deff3}
< B^*(p',\epsilon ' )\vert V^\mu \vert B^*(p,
\epsilon)>=F_1(q^2)(\epsilon\cdot \epsilon ')P_\mu+F_2(q^2)[
\epsilon_\mu (\epsilon'\cdot P) +\epsilon '_\mu (\epsilon\cdot P)]\nonumber\\
+F_3(q^2){(\epsilon\cdot P) (\epsilon'\cdot P) \over m_{B^*}^2}P_\mu+
F_4(q^2)[\epsilon_\alpha (\epsilon'\cdot P)-\epsilon'_\alpha
 (\epsilon\cdot P)]
{g_{\mu\alpha} q^2-q_\mu q_\alpha \over m_{B^*}^2}\,,
\end{eqnarray}
  where $V^\mu=\bar b\gamma^\mu b$,  $\epsilon (\epsilon ')$ denote
 the polarization vectors of the $B^*$ mesons, $P=p+p'$ and $q=p-p'$.

The form factors defined above have cuts in the complex plane 
$t=q^2$, from the threshold $t_0$ for $B\bar B$ production to
infinity. The effect of the lower branch cuts due to light
intermediate states ($\pi\pi, KK$, etc) is negligible \cite{raf}.
The three resonances $\Upsilon(1S)$, $\Upsilon(2S)$, $\Upsilon(3S)$
 with masses lower than $2m_B$ produce
additional singularities, which
 can be approximated by poles 
on the real axis below $t_0$ \cite{raf1}.
 On the other hand, heavy quark symmetry predicts 
definite relations among the form factors (\ref{deff1})-(\ref{deff3}) near
the zero recoil point $w=1$ ($w=v\cdot v'$, $v$ and $v'$
being the velocities of the initial and final meson, respectively). 
In this region some of the form factors in  (\ref{deff2}) and (\ref{deff3})
are approximately equal to the elastic form factor (\ref{deff1}) and
other vanish. Specifically, for $w\approx 1$ one has
\begin{equation}\label{iw}
V(w)=-F_1(w)=F_2(w)=F(w) \, ,~~ F_3(w)=F_4(w)=0\, ,
\end{equation}
and we recall that $F(w)$ satisfies the normalization condition
\begin{equation}\label{norm}
F(1)=1\, .
\end{equation}
We are interested in finding restrictions on the slope of this function
at zero recoil, or
the so called charge radius, defined as
\begin{equation}\label{radius}
\rho^2=-F'(1),
\end{equation}
which differs by ${16\over 75}{\rm log}\alpha_s(m_b)$ \cite{falk} from the charge radius
$\rho^2_{IW}$ of
the universal Isgur-Wise function \cite{isgur}.

As in the derivation of the usual QCD sum sules, for studying
the form factors (\ref{deff1}-\ref{deff3}) we start by considering
the vacuum polarization tensor due to the current 
$V^\mu$:
\begin{equation}\label{polar}
\Pi^{\mu \nu}(q)=(q^\mu q^\nu-g^{\mu \nu}q^2)\Pi(q^2)=i
\int dx e^{iqx}<0\vert T(V^\mu (x)V^\nu (0))\vert 0>.
\end{equation}
The first derivative of 
the invariant amplitude  $\Pi(q^2)$ satisfies the dispersion relation
\begin{equation}\label{disprel}
\Pi'(q^2)={1\over \pi}\int_{0}^\infty{{\rm Im}\Pi (t)\over (t-q^2)^2}dt
\,,
\end{equation}
the spectral function being defined by the unitarity relation
\begin{eqnarray}\label{unit}
(q^\mu q^\nu-g^{\mu \nu}q^2){\rm Im} \Pi(t+i\epsilon)~= {1\over 2}
\sum_\Gamma \int d\rho_\Gamma (2\pi)^4\delta^{(4)}(q-p_\Gamma )\nonumber\\
\times<0\vert
V^\mu (0)\vert \Gamma ><\Gamma \vert V^\nu (0)^{+}\vert 0>.
\end{eqnarray}
Here the summation is over all possible hadron states $\Gamma$ with
appropriate flavor quantum numbers, with an integral over the phase space
allowed to each intermediate state. 
We shall include in this sum 
the three $\Upsilon$ resonances with masses lower than the threshold
of the $B\bar B$ production and the contribution
 of the two-particle states $\vert B \bar B>, 
 \vert B \bar B^*+ B^* \bar B>$
and $\vert B^* \bar B^*>$ above this threshold (the 
$\Upsilon(4S)$ resonance is not included, 
in order to avoid double counting \cite{raf1}). This
contribution can be evaluated in a straightforward way, by
using the definitions (\ref{deff1}-\ref{deff3}) of the form factors,
 performing the phase-space integration and the summation over the
polarizations of the $B^*$ intermediate states. Taking into account
the positivity of the spectral function of $\Pi$, which follows
from (\ref{unit}) we obtain the following inequality:
\begin{eqnarray}\label{inegal}
{1\over \pi}{\rm Im}\Pi(t+i\epsilon)\geq {27\over 4\pi\alpha^2} 
\sum_i M_{\Upsilon_i}
 \Gamma_{\Upsilon_i} \delta(t-M_{\Upsilon_i}^2) \nonumber \\
 +{n_f\over 48\pi}\left\{\left(1-{t_0\over t}\right)^{3/2} 
\vert F(t) \vert ^2 \theta (t-t_0)\right.
+ (1-{t_0^*\over t})^{3/2}(1-{t_1^*\over t})^{3/2}{2t\over t_0^*}
\vert V(t) \vert^2 \theta (t-t_0^*) \nonumber \\
\left. +(1-{t_0^{**}\over t})^{3/2}~\left[2 \vert F_1(t) \vert^2 
+{4t\over t_0^{**}}\vert F_2(t) \vert^2 + \vert \widehat{F}_3(t)
 \vert^2 
+({4t\over t_0^{**}})^2\vert F_4(t)\vert^2\right]
\theta (t-t_0^{**})\right\}\,,
\end{eqnarray}
  where 
\begin{equation}\label{f3bar}
 \widehat{F}_3(t) = ({2t\over t_0^{**}}-1)F_1(t)+
{2t\over t_0^{**}}F_2(t)+{2t\over t_0^{**}}({2t\over t_0^{**}}-1)F_3(t).
\end{equation}
In (\ref{inegal}) the widths $\Gamma_{\Upsilon_i}$ are defined through the parametrization
\begin{equation}\label{sigmaee}
\sigma(e^+e^-\rightarrow\Upsilon_i)=12\pi^2\delta(t-M^2_{\Upsilon_i})
{\Gamma_{\Upsilon_i}\over M_{\Upsilon_i}},
\end{equation}
of the cross section for $\Upsilon$ production,
 $t_0=4m_B^2,~
t_0^*=(m_B+m_{B^*})^2$ and
$t_0^{**}=4m_{B^*}^2$ are the thresholds for $B\bar B$,
$B\bar B^*$ and $B^*\bar B^*$ production, respectively.
We used the notation $t_1=(m_{B^*}-m_{B})^2$ and
$n_f=3$ is the number of light quark flavors which give identical
contribution in the unitarity sum \cite{raf1}. It was convenient to 
write the contribution of the last four
form factors in a "diagonal" form, as a sum of moduli squared of
functions with the same analyticity properties as the original form
factors,
which will allow the application of the mathematical technique presented below.
This calculation was rather tedious and for simplicity
 these terms were omitted in the previous paper \cite{capspin}
devoted to this problem.

By combining the dispersion relation (\ref{disprel}) with the 
unitarity inequality (\ref{inegal}) we obtain the following
integral condition for the form factors of interest:
\begin{eqnarray}\label{condit} 
\widetilde \Pi'(q^2)\geq {n_f\over 48\pi^2}\left\{\int_{t_0}^\infty{1\over (t-q^2)^2}
\left(1-{t_0\over t}\right)^{3/2} 
\vert F(t) \vert ^2 dt\right.\nonumber \\
+\int_{t_0^*}^\infty {1\over (t-q^2)^2}
(1-{t_0^*\over t})^{3/2}(1-{t_1^*\over t})^{3/2}{2t\over t_0^*}
\vert V(t) \vert^2 dt 
+\int_{t_0^{**}}^\infty {1\over (t-q^2)^2}
(1-{t_0^{**}\over t})^{3/2}~\nonumber \\
\times\left. \left[2 \vert F_1(t) \vert^2 
+{4t\over t_0^{**}}\vert F_2(t) \vert^2 
 + \vert \widehat{F}_3(t) \vert^2 
+({4t\over t_0^{**}})^2\vert F_4(t)\vert^2\right]dt\right\},
\end{eqnarray}
where
\begin{equation}\label{pitilde}
\widetilde\Pi'(q^2)= \Pi'(q^2)- {27\over 4\pi\alpha^2 } 
\sum_i { M_{\Upsilon_i}
 \Gamma_{\Upsilon_i} \over (q^2-M_{\Upsilon_i}^2)^2}. 
\end{equation}
In the euclidian region $q^2<0$ the function $\Pi'(q^2)$
 can be calculated by applying QCD
 perturbation theory, with nonperturbative corrections included by
means of OPE. Due 
 to the large value of $m_b$, the QCD expression of $\Pi'(q^2)$
can be used also at $q^2=0$ or even at positive $q^2$ much less than $4m_b^2$.
Moreover, in this case the nonperturbative 
corrections are shown to be entirely negligible 
\cite{svz}.
In  the previous works \cite{raf}-\cite{boyd} only the lowest order
(one-loop) perturbative polarization function was used as input in
eq.(\ref{pitilde}) (the terms containing the $\Upsilon$ 
poles being also omitted). In the present analysis
we introduce explicitely in (\ref{pitilde}) the contribution of the 
$\Upsilon$ resonances, using the experimental information on 
$\Gamma_{\Upsilon_i}$. In the same time
 we use as input the expression of the polarization function up
to three loops \cite{svz}-\cite{kuhn}:
\begin{equation}\label{expans}
\Pi'(q^2)=\Pi'^{(0)}(q^2)+{\alpha_s (\mu^2)\over \pi}\Pi'^{(1)}(q^2)+
\left({\alpha_s(\mu^2)\over \pi}\right)^2\Pi'^{(2)}(q^2,\mu^2)
\end{equation}
with the $\overline{MS}$ coupling $\alpha_s (\mu^2)$ defined in the
conventional way.
We use the standard expressions \cite{svz} 
\begin{equation}\label{pi01}
\Pi'^{(0)}(q^2)={1\over 32\pi^2 m_b^2}\int_0^1{v(3-v^2)
\over (1-q^2x/4m_b^2)^2}dx,
\end{equation}
\begin{equation}\label{pi02}
\Pi'^{(1)}(q^2)={1\over 24\pi m_b^2}\int_0^1{v(3-v^2)
\over (1-q^2x/4m_b^2)^2}\left [{\pi\over 2v}-{v+3\over 4}\left ({\pi\over
2}-{3\over 4\pi}\right )\right] dx,
\end{equation}
with $v=\sqrt{1-x}$. As concerns the last term in (\ref{expans}),
 we shall use the Taylor series around $q^2=0$:
\begin{equation}\label{pi03}
\Pi'^{(2)}(q^2,\mu^2)={3\over 64\pi^2 m_b^2}\sum_n nC_n
\left({q^2\over 4m_b^2}\right)^{n-1},
\end{equation}
the coefficients $C_n$ being given in eq.(11) of ref.
\cite{kuhn} (we recall that they depend explicitely on the
normalization scale $\mu$). 

With the above expressions, the input entering (\ref{condit}) is
completely specified and this inequality can be viewed as an integral
quadratic condition for the form factors of interest along the 
unitarity cut. 
By applying standard techniques of analytic functions \cite{duren},
extended to "vector-valued functions" (see \cite{capcomp} and the references
therein) we shall obtain from this condition  a quadratic inequality
 relating the values of the form factors and
their derivatives at the zero recoil point.
 Using then the relations (\ref{iw}-\ref{radius})
we shall finally express the derived inequality as a constraint
 on the charge radius  (\ref{radius}). 

We first  conformally map the cut $t=q^2$
plane onto the unit disk in the complex plane $z$, 
 such that the unitarity cut becomes
the boundary $\vert z \vert = 1$. Actually,
since the integrals appearing in (\ref{condit}) have different thresholds,
 we shall use for them different conformal mappings 
. More precisely, we take
\begin{equation}\label{mapp}
z(t)~=~{\sqrt {t_0-t}-\sqrt{t_0}\over \sqrt {t_0-t}+\sqrt{t_0}}
\end{equation}
for the first integral in the right hand side of
 (\ref{condit}) and similar expressions, with 
$t_0$ replaced
by $t_0^*$ and $t_0^{**}$, respectively,
 for the second and the third integral.
By the mapping (\ref{mapp}) the threshold $t_0$ becomes 
 $z=-1$ and the zero recoil point $w=1$ (equivalent to $t=0$,
since $ w=1-{t\over 2m_B^2}$) is applied onto the origin $z=0$.
Similarly, using the mappings  suitable
for the other integrals
in (\ref{condit}) as explained above, 
the thresholds $t_0^*$ and $t_0^{**}$  become also $z=-1$ and 
the corresponding zero recoil
point is applied on the origin. It is easy to see that the conformal
mappings  used for the second and the third integrals transform
 the threshold $t_0$ into a point situated inside the unit circle, 
 close to $-1$.
By performing the above changes of variable,
 all the integrals in (\ref{condit})
become integrals along the same contour, i.e. the boundary $z=e^{i\theta}$
of the unit disk. 

It is convenient to introduce a compact notation by defining the
following functions of the variable $z$:
\begin{eqnarray}\label{f}
f_1(z)=F(t),~f_2(z)=V(t),\nonumber \\
~f_3(z)=F_1(t),~f_4(z)=F_2(t),~f_5(z)=\widehat{F}_3(t),~f_6(z)=F_4(t),
\end{eqnarray}
where $\widehat{F}_3$ is defined in (\ref{f3bar}). Using the conformal
mappings defined above, the normalization
condition (\ref{norm}) and the definition (\ref{radius}) of the charge
radius, one can show easily that the functions $f_i(z)$ satisfy the relations
\begin{equation}\label{forig}
f_i(0)=1,~f'_i(0)=-8\rho^2, ~i=1,..5;~~~f_6(0)=0,
\end{equation}
the derivative being with respect with $z$.
 Moreover, following the
standard technique presented in \cite{raf}-\cite{capzeit},\cite{meiman}
 we shall
 define a set of functions  $\phi_i(z)$ analytic and without zeros 
in the unit disk, whose moduli
squared on the boundary are proportional to the positive weights appearing
in the integrals (\ref{condit}), multiplied by the Jacobian
 $\vert {dt\over dz}\vert$ of the conformal mapping (\ref{mapp}). These
functions can be constructed easily and unambigously, by using the
relations
\begin{eqnarray}\label{aux}
t={4t_0\over (1-z)^2}, ~~~~~
\left(1-{t_0\over t}\right)^{3/2}={(1+z)^3\over 8}, \nonumber\\
{dt\over dz}=4t_0{1+z\over (1-z)^3},~~~~~
{1\over (t-q^2)}=\left({2\over 1-d}\right)^2{1\over t_0}
{(1-z)^2\over (1-zd)^2},
\end{eqnarray}
which follow from (\ref{mapp}), with
\begin{equation}\label{d}
d~=~{\sqrt {t_0-q^2}-\sqrt{t_0}\over \sqrt {t_0-q^2}+\sqrt{t_0}}.
\end{equation}

With these defintions, we can write the inequality (\ref{condit}) in the
equivalent form
\begin{equation}\label{l2norm}
{1\over 2\pi} \int_0^{2\pi}\sum_{i=1}^6 \vert \phi_i(\theta) f_i(\theta)
\vert ^2 d\theta\leq 1\, 
\end{equation}
where the functions $\phi_i(z)$, obtained
using (\ref{aux}) can be written in a compact form as
\begin{equation}\label{phi}
\phi_i (z)=\phi_i (0){(1+z)^{a_i} (1-z)^{b_i}\over (1-zd_i)^{c_i}}.
\end{equation}
The parameters entering this expression are as follows:
\begin{eqnarray}\label{phi0}
 \phi_1 (0)={(1-d)^2\over 32m_B}\sqrt{{ n_f\over 6\pi\widetilde \Pi'(q^2)}}, ~~
a_1=2,~b_1=1/2,~c_1=2,~d_1=d\nonumber \\
 \phi_2 (0)=2\sqrt{2}\phi_1(0),~~ a_2=2,~b_2=-3/2,~c_2=2,~d_2=d\nonumber \\
 \phi_3 (0)={(1-d^*)^2\over 32m_{B^*}}\sqrt{{ n_f\over 3\pi\widetilde \Pi'(q^2)}},
~~a_3=2,~b_3=1/2,~c_3=2,~d_3=d^*\nonumber \\
 \phi_4 (0)=2\sqrt{2}\phi_3(0),~~a_4=2,~b_4=-3/2,~c_4=2,~d_4=d^*\nonumber \\
 \phi_5 (0)={\phi_3(0)\over \sqrt{2}},~~a_5=2,~b_5=1/2,~c_5=2,
~d_5=d^*\nonumber \\
 \phi_6 (0)=8\sqrt{2}\phi_3(0),~~a_6=2,~b_6=-3/2,~c_6=2,~d_6=d^*\, ,
\end{eqnarray}
with $\widetilde \Pi'$ defined in (\ref{pitilde}) and 
$d$ in (\ref{d}) ($d^*$ is obtained from $d$ by replacing $m_B$ by
$m_{B^*}$).

As discussed above, the form factors appearing in (\ref{condit})
have three poles on the real axis below the threshold $t_0=4m_B^2$, 
due to the three
$b\bar b$ bound states $\Upsilon(1S), \Upsilon(2S)$ and
$\Upsilon(3S)$ with masses smaller than the threshold for
 $B\bar B$ production. The positions of these
poles are known from the experimental masses of the $\Upsilon$
resonances, but the residues are unknown,
containing the unphysical $\Upsilon B\bar B$ or $\Upsilon B\bar B^*$ couplings
 \cite{raf1}.
The form factors $V$ and $F_i$ have in addition
branch points at the threshold $t_0$ of the $B\bar B$ production, below
the beginning of the corresponding unitarity cut.
If an estimate of the discontinuity across these cuts
were available, the treatment of these subthreshold singularities
in  the present formalism could be done exactly \cite{capcut}
(the method was applied recently in \cite{irmatt} to the $B\rightarrow D$
 form factors). In what follows we shall resort to a pole
approximation, keeping  only the contribution
of the $\Upsilon$ resonances situated below the thresholds $t_0^*$ and
$t_0^{**}$, respectively.
Using  $m_B=$5.279 GeV, $m_{B^*}=$5.324 GeV
and the masses  of the  $\Upsilon$
resonances ($M_{\Upsilon_1}=$9.460 GeV, $M_{\Upsilon_2}=$ 10.023 GeV,
$M_{\Upsilon_3}=$ 10.355 and $M_{\Upsilon_3}=$10.580 GeV) one
can easily see that the form factor $V(t)$ has only three poles below
its unitarity threshold, much like $F(t)$, while $F_i(t)$ have four poles.
Passing to the functions $f_i(z)$ according to (\ref{f}) and using
 the conformal transformation (\ref{mapp}), we find
that the functions $f_1(z)$ and $f_2(z)$  have inside $\vert z\vert< 1$
three poles situated at the points
\begin{equation}
z_1=-0.38\, , z_2=-0.52\, , z_3=-0.67.
\end{equation}
We neglected here the difference between $m_B$ and $m_{B^*}$,
which is entirely justified as long as the singularities remain
the same.
 As concerns the remaining functions $f_i, i\geq 3$, they have
four poles, with positions
\begin{equation}
z_1^*=-0.37\, , z_2^*=-0.49\, , z_3^*=-0.62\, ,z_4^*=-0.79,
\end{equation}
obtained by using the conformal mapping (\ref{mapp}) with $t_0$
replaced by $t_0^{**}$ and $t$ by  $M_{\Upsilon_i}$.

The inequality (\ref{l2norm}) has the form of an $L^2$ norm
condition \cite{duren} involving several functions. We derive
from it constraints on the functions $f_i$ and their
derivatives at the origin $z=0$, which corresponds through the conformal
mapping to the point of zero recoil $w=1$. If the functions $f_i$ were analytic,
this would be very easily done,  by applying standard techniques in the
Hilbert space $H^2$ \cite{duren}. However, as shown above, the functions have
a finite number of poles, with known positions but unknown residua. The
simplest treatment of this situation is based on the
 technique of Blaschke functions \cite{duren} (the method was applied
previously in \cite{capzeit}-\cite{boyd}).
 We define the following  functions
\begin{equation}\label{blas}
B(z)=\prod_{k=1}^3 {(z-z_k)\over (1-zz_k)},~~
B^*(z)=\prod_{k=1}^4 {(z-z_k^*)\over (1-zz_k^*)}
\end{equation}
where we took into account that $z_k$ and $z_k^*$ are real.

As seen from (\ref{blas}) the functions $B(z)$ and $B^*(z)$ have 
modulus equal to 1 on the
boundary of the unit disk (i.e. for $z=e^{i\theta}$). Therefore,
 we can insert the modulus squared of the function 
$B(\theta)$ (or $B^*(\theta)$) in the
integral appearing in (\ref{l2norm}), without spoiling the inequality or losing
information. The relation (\ref{l2norm}) is thus equivalent to
\begin{equation}\label{l2an}
{1\over 2\pi} \int_0^{2\pi}\sum_{i=1}^6 \vert \phi_i(\theta) B_i(\theta)
f_i(\theta)\vert ^2 d\theta\leq 1,
\end{equation}
where we denoted
\begin{eqnarray}\label{bi}
B_i(z)=B(z)~~ (i=1,2)~;~
B_i(z)=B^*_i(z)~~ (i=3,6).
\end{eqnarray}
 But the products $B_i(z)f_i(z)$  are functions analytic in
$\vert z \vert < 1$, the poles of the form factors $f_i$ being
compensated by the zeros of the  functions $B_i(z)$.
 We can apply therefore the
well-known results of interpolation theory for vector-valued analytic
functions (see \cite{capcomp} and references therein)
to obtain from (\ref{l2an}) constraints on the form factors at points
inside the analyticity domain. 
In particular, being interested in finding bounds on the charge radius 
(\ref{radius}) which appear in (\ref{forig}),
we shall apply an inequality of the 
 Schur-Caratheodory type \cite{duren} at $z=0$:
\begin{equation}\label{sc}
\sum_{i=1}^6\left[\phi_iB_if_i)^2(0)+(\phi_iB_if_i)'^2(0)\right]\leq
1\, . \end{equation}
It is important to note that up to now the form factors $f_i$ 
were treated as independent functions, without assuming that they coincide
along the unitarity integrals.
We use now heavy quark spin symmetry, which imply the relations (\ref{forig}).
Then (\ref{sc}) can be written as an inequality for the charge radius
\begin{equation}\label{scmod}
\sum_{i=1}^5\phi_i^2(0)B_i^2(0)+\sum_{i=1}^5\left[\phi_i(0)B_i'(0)+\phi_i'(0)
B_i(0)-8\rho^2\phi_i(0)B_i(0)\right]^2\leq 1\, .
\end{equation}
 The function $f_6$ does not contribute, due to the last condition
 in (\ref{forig}). The inequality (\ref{scmod}) can be written as
\begin{equation}\label{eq}
a(\rho^2)^2-2b\rho^2+c\leq 0,
\end{equation}
where
\begin{eqnarray}\label{abc}
a=64\sum_{i=1}^5 B^2_i(0)\phi^2_i(0)\nonumber \\
b=8\sum_{i=1}^5 B_i(0)\phi_i(0)[\phi_i'(0) B_i(0)+\phi_i(0) B_i'(0)]
\nonumber \\
c=\sum_{i=1}^5 [\phi_i'(0) B_i(0)+\phi_i(0) B_i'(0)]^2
+\sum_{i=1}^5 B^2_i(0)\phi^2_i(0) -1.
\end{eqnarray}
The quantities $\phi_i(0)$, $\phi_i'(0)$, $B_i(0)$ and $B_i'(0)$,
entering the above coefficients, are
calculable from the relations (\ref{phi}),(\ref{phi0}) and (\ref{blas})
 and contain all the dynamical information in the problem.
\section{Results and conclusions}
We discuss now the lower and upper bounds on the
charge radius $\rho^2$ calculated from the above equation (\ref{eq}).
First we recall that the results previously reported in 
\cite{raf1} and the second reference \cite{capzeit} can be obtained
by restricting the sums in the expressions (\ref{abc}) to a single term, $i=1$.
In the above works only the lowest order term $\Pi'^0$ in the expansion 
(\ref{expans}) of $\Pi'$ was retained and the contribution of the $\Upsilon$
poles in the relation (\ref{pitilde}) was dropped out.  Also,
for simplicity  the choice $m_b=m_B$ for the mass of the $b$ quark was made,
and the value of $q^2$ which enters as a parameter in 
eq. (\ref{condit}) was taken $q^2=0$.
With these restrictions, eq. (\ref{eq}) gives the interval
$-4.5\leq \rho^2\leq 6.1$
 already reported in \cite{raf1}. Keeping two
terms $(i=1,2)$ in the sums appearing in (\ref{abc}), with the 
same numerical input as just
described, we recover the interval $-0.9\leq \rho^2\leq 2.60$ obtained
in \cite{capspin}. Finally, with all the five terms in the sums, i.e. by
including all the form factors of the ground states $B$ and $B^*$, we obtain
with the same input the range $-0.35\leq \rho^2\leq 2.15$. This
result shows the improvement which can be obtained by fully exploiting
 spin symmetry in the ground state $B$ doublet.

As we mentioned, the above results were obtained with
 some simplifying assumptions concerning the input.
It is therefore of interest to perform  the analysis with a more
realistic input, according
to the complete formulas given above.
The main improvement is the QCD expression (\ref{expans}) of the polarization
function up to three loops corrections. 
This expression depends on the scale $\mu$ which appears in the 
$\overline{MS}$ coupling $\alpha_s(\mu)$ and in
the coefficients $C_n$ of the Taylor expansion (\ref{pi03}). We 
shall use in our analysis two scales, namely $\mu=m_b$, for which
the coefficients $C_n$  are \cite{kuhn}
\begin{equation}\label{coef1}
C_1=32.73,~~C_2=33.24,~~C_3=29.61,~~C_4=26.94,
\end{equation}
and $\mu=2m_b$, which gives 
\begin{equation}\label{coef2}
C_1=49.57,~~C_2=43.31,~~C_3=37.91,~~C_4=33.92.
\end{equation}
We note that  for the above choices of $\mu$ the coefficients $C_n$
do not depend on the specific value of $m_b$.
Although the coefficients in (\ref{coef1}) and {\ref{coef2}) 
are quite different,  the final results,
i.e. the bounds on $\rho^2$, turn out to be practically the same.

The expressions given in (\ref{expans}-\ref{pi03}) 
were obtained using on shell renormalization,
which means that $m_b$ is the pole mass. In the present work we shall
treat this mass as a parameter in the reasonable range 
$4.7{\rm GeV}-5.{\rm GeV}$. 
For these values of $m_b$ and the choices of $\mu$ made above, the 
two-loops correction in the expansion (\ref{expans}) for $q^2=0$ represents 
about $30\%$ of the lowest order term, while the contribution of
the three-loops diagrams
is of about $10\%$ (we used $\alpha_s(5.{\rm GeV})=0.21$ and
$\alpha_s(10.{\rm GeV})=0.18$ \cite{data}). As we already pointed out,
 for heavy quarks one can extend the
validity of the QCD perturbative expansion of the polarization 
function even at positive
values of $q^2$, below the threshold $t_0$. As an
exemple, 
for $q^2=50~ {\rm GeV}^2$ the two-loops term represents a correction of
 about $45\%$ of the one loop term, while the three loops contribute
in addition with approximately $20\%$. In the present formalism better
 results, i.e. stronger bounds on $\rho^2$, are obtained for larger $q^2$.
On the other hand, the  increased
contribution of the higher order QCD corrections for the polarization
function prevents us taking $q^2$ too close to the hadronic singularities.
 We shall
take in what follows $q^2$ in the range $0-50 {\rm GeV}^2$, noticing
 that the relative magnitude of the perturbative
corrections does not dramatically change in this domain.

We recall that much smaller values for
 the QCD perturbative
corrections to the polarization function of heavy quarks were 
reported in \cite {svz} (see also \cite{rry}). The idea applied in
these works was to express
 the pole mass $m_b$ in (\ref{pi01}) and (\ref{pi02}) in terms of
 an euclidian mass defined to first order in $\alpha_s$.
This had the effect of reducing the procentual contribution of
the two-loop correction, especially in the
high order derivatives of the function $\Pi(q^2)$, of interest 
in the QCD sum rules for heavy quarks. The recent calculation of the
polarization function up to three loops \cite{kuhn} 
allowed us to use a more exact expression of $\Pi'$, without
resorting to the rather arbitary procedure adopted in \cite{svz}.

The contribution of the $\Upsilon$ poles in the expression
 (\ref{pitilde}) was evaluated using
the numerical values $\Gamma_{\Upsilon_1}=1.34~ {\rm keV}$, 
$\Gamma_{\Upsilon_2}=0.56~ {\rm keV}$ and 
$\Gamma_{\Upsilon_1}=0.44~ {\rm keV}$ \cite{data}. 
The poles bring a positive
contribution to the spectral function according to
(\ref{inegal}) and their inclusion 
 improves the bounds in a significant way.

In Fig.1 we present the upper and lower
bounds on the charge radius $\rho^2$ 
of the $B$ meson elastic form factor, computed from (\ref{eq}),
 with the input described above,
for $m_b$ in the range $4.7{\rm GeV}-5.{\rm GeV}$.
As we mentioned, the two choices of the scale $\mu$ adopted above
give almost identical results. The solid curve corresponds
to the choice $q^2=0$, the dashed one to $q^2=50~{\rm GeV}^2$. 
Taking larger values of $q^2$ we obtain much stronger bounds, 
 but inconsistencies appear around $60.{\rm GeV}^2$ (the pole contribution
exceeds the QCD expression of $\Pi'(q^2)$, signaling that 
 a better estimation of the input is necessary). As seen from Fig.1,
the predictions for the charge radius are rather sensitive to
the value of the pole mass $m_b$, larger values of the mass leading to 
stronger bounds.

The upper and lower bounds given in Fig.1 represent
 the best results that can be derived, using
a realistic input and fully exploiting  heavy quark spin symmetry
for the ground state $B$ and $B^*$ mesons.
We recall that the present derivation was possible by resorting to a
 a more powerful technique of analytic
functions, which allowed the simultaneous treatment of several form
factors as independent functions. The specific unitarity thresholds of
the different form factors and their subthreshold singularities were
correctly taken into account.
 Heavy quark spin symmetry was invoked finally by assuming that
various 
form factors coincide near the zero recoil point,  which
is entirely legitimate.

The technique applied in this paper can be easily generalized
(see \cite{capzeit}, \cite{irmatt}) 
to include higher derivatives of the form factors at the zero recoil
point. In this way, for instance, quite strong correlations among
 the slope and the convexity of the elastic form factor $F(t)$ can 
be derived. A second, more interesting generalization is to
 include in the unitarity sum the contribution of 
the excited states ($B^{**}$) with orbital momentum
$L=1$. By applying the techniques
used in this work, it is possible to derive an
 inequality connecting the form factors of the ground states  $B$ and $B^*$
to the transition 
form factors between $B^{**}$  and the ground states.
A new sum rule for these form factors, 
 similar to  the well-known inequalities of Bjorken
\cite{bjorken} and Voloshin \cite{volos},
 will be reported in a future paper \cite{future}.
\vskip0.5cm
\noindent
{\Large\bf Aknowledgment}
\vskip0.5cm
\noindent
One of the authors (I.C.) is pleased to aknowledge interesting discussions
on topics related to the present work with
Patricia Ball, Laurent Lellouch and Matthias Neubert.

\vskip0.5cm
\noindent
{\Large\bf Figure caption}
\vskip0.3cm
\noindent
FIG.1: Upper and lower bounds on the charge radius of the elastic
form factor of the $B$ meson for various values of the pole mass $m_b$.
 The solid line corresponds to  
$q^2=0$, the dashed one to $q^2=50~ {\rm GeV}^2$. 

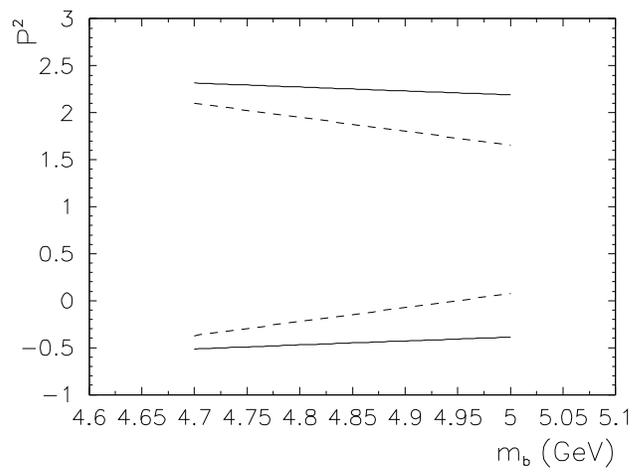
\end{document}